\documentclass{superfri}
\usepackage{graphicx}
\usepackage[hidelinks]{hyperref}

\usepackage{listings}

\usepackage{xcolor}
\usepackage{amsmath}
\usepackage[T2A]{fontenc}
\usepackage[utf8]{inputenc}
\usepackage{multirow,tabularx}

\begin{document}

\author{Alexander~V.~Smirnov,\footnote{\label{msu}Lomonosov Moscow State University, Moscow, Russian Federation}\orcidID{0000-0002-7779-6735}
\and
Boris~I.Rozhnov$^{1}$\orcidID{0009-0001-3313-8001}
\and
Vadim~V.~Voevodin$^{1}$\orcidID{0000-0003-1897-1828}}

\title{GPU Implementation of Zippel Method for Feynman Integral Reconstruction}

\maketitle{}

\begin{abstract}%
The Zippel algorithm performs a rational reconstruction of multivariate polynomials and aims specifically at the sparse case. It is applied in different fields of science, lately becoming an important step in Feynman integral reduction in elementary particle physics. In some cases with multiple variables it might become a bottleneck for the whole evaluation so that different optimizations are required. In this paper we describe how we ported the classical Zippel algorithm together with its balanced version for rational functions to graphical processor units and perform its evaluation on several GPUs.

\end{abstract}

\section*{Introduction}
\label{sec:intro}

The main motivation for this paper lies in the field of elementary particle physics, namely Feynman integral reduction \cite{Chetyrkin:1981qh}, one of the key steps of evaluation of Feynman integrals. However a Feynman integral reduction technically means solving a huge sparse linear system with coefficients being rational functions of multiple variables; hence, the scope of the paper and possible applications are much more broad.

The classical approach to Feynman integral reduction was to solve the system directly on a large enough server \cite{Laporta:2000dsw, Anastasiou:2004vj, Lee:2012cn, Lee:2013mka, Maierhofer:2017gsa, Smirnov:2013dia, Smirnov:2014hma, vonManteuffel:2012np,  Maierhofer:2018gpa, Smirnov:2019qkx, Klappert:2020nbg, lange2025kira3integralreduction}, but with increasing complexity and the availability of a supercomputer infrastructure, new methods were required. Therefore, appeared an approach with a rational reconstruction of functions treating the unknown coefficients as black-box rational functions of multiple variables that have to be reconstructed afterwards \cite{vonManteuffel:2014ixa,Peraro:2016wsq,Peraro:2019svx,Klappert:2019emp,Laurentis:2019bjh,DeLaurentis:2022otd,Magerya:2022hvj,Belitsky:2023qho}. Within this ``modular'' approach the reduction is first performed multiple times with fixed values of variables in modular fields over large prime numbers (fitting into $2^64$ to utilize machine-size arithmetics), and then the functions are reconstructed. One should not confuse reconstruction and interpolation; knowing the values of a rational function in multiple points one surely has an unlimited number of rational functions with such properties, but the reconstruction methods are aiming to find such a function that the following probes (evaluations of the function at other points) also match the guessed function.

The reconstruction method has a long history starting with Newton interpolation reconstructing a polynomial of one variable.

\begin{align}
\label{eq:newton}
f_N (x) 
&= {\rm Newton}_x [f (x), N]
\\
&
\equiv 
a_1 + (x - x_1) \Big[ a_2 + (x - x_2) \big[ a_3 + (x - x_3) \left[ a_4 + \dots \right] \big] \Big]
\, . \nonumber
\end{align}

For multivariate polynomial reconstruction one can proceed variable by variable, i.e. when reconstructing from $n$ variables to $n+1$ variables one takes a needed number of reconstructed functions $f(x_1, x_2, \ldots, x_n, x_{n+1, i})$ for different values of $x_{n+1, i}$ and runs the univariate Newton reconstruction.

For univariate rational functions there is the Thiele reconstruction

\begin{align}
\label{eq:thiele}
f_T (x) 
&= {\rm Thiele}_x [f (x), T]
\\
& 
\equiv
b_0 + (x - x_1) \left[ b_1 + (x - x_2) \left[ b_2 + (x - x_3)\left[ b_4 + \dots \right]^{-1} \right]^{-1}\right]^{-1}
\, . \nonumber
\end{align}

The situation becomes more complicated with multivariate rational functions since the recursive Thiele formula is a combination of continued fractions and is technically too complex to be evaluated in real examples. Therefore, for multivariate rational functions, other methods are used, one of those being the balanced Newton reconstruction proposed in \cite{Belitsky:2023qho} and the balanced Zippel reconstruction proposed in \cite{Smirnov:2024onl} for sparse multivariate functions.

The balanced Zippel reconstruction of rational functions is the main subject of this paper. It is based on the Zippel reconstruction of polynomials suggested initially in \cite{10.1007/3-540-09519-5_73} and developed in \cite{10.1145/62212.62241, 10.1145/345542.345629}. The Zippel method is an approach with the lowest complexity for the sparse polynomial reconstruction. It is already widely applied in programs for Feynman integral reduction using the modular approach with the following reconstruction. However, for some cases, the Zippel reconstruction itself might become a bottleneck for the whole approach due to its quadratic complexity by the number of terms in the skeleton polynomial (see next section for details) which might be measured in millions. Hence any possible optimization of the Zippel algorithm is needed.

There is a number of public Zippel algorithm implementations, mostly in some open-source repositories. Speaking of its applications to Feynman algorithms, programs such as  \texttt{Kira} or \texttt{FIRE} use the Zippel algorithm. 

The main contribution of this paper is an implementation of the Zippel algorithm on GPUs (Graphical Processing Units). According to our information, this is the first such implementation of this algorithm on GPUs. Currently we publish a description and evaluation results, while the code is in a private repository related to Feynman integrals, but we intend to make it public as a part of the new \texttt{FIRE} version this year. We are also considering a possible separation of the modular GPU and reconstruction part providing it as a separate library.

The rest of the paper is organized as follows. Section~\ref{sec:zippel} describes the original Zippel algorithm. Section~\ref{sec:impl} is devoted to our implementation of this algorithm on GPUs. Section~\ref{sec:eval} provides evaluation results and benchmarks.

\section{Description of Zippel Method}
\label{sec:zippel}

The Zippel method is a reconstruction procedure from a polynomial with $k-1$ variables to a polynomial with $k$ variables. The traditional approach for dense polynomials (meaning that if one considers all possible monomials of a given degree most of the coefficients are non-zero) is to proceed with a Newton reconstruction formula. However for sparse polynomials this is quite inefficient because it requires a large number of probes for the reconstruction. 

Let us remind the method taking a few formulas from our paper \cite{Smirnov:2024onl}. Suppose that we have already reconstructed a polynomial $f(x_1, \ldots, x_{k-1}, c_0)$, where $c_0$ is some constant value of $x_k$. There might be more variables, but they should be fixed at this point. We aim to run the Newton reconstruction for $x_k$, so we need similar reconstructed polynomials for other values of $x_k$.

Suppose we take another value $c_i$ of $x_k$. We consider the existing polynomial $f(x_1, \ldots, x_{k-1}, c_0, \ldots)$ as a skeleton, take all its non-zero monomials and assume that the set of nonzero monomials will remain the same for the yet unknown $f(x_1, \ldots, x_{k-1}, c_i, \ldots)$ so that it can have the following form:

\begin{equation}
f(x_1, \ldots, x_{k-1}, c_i, \ldots) =  a_1 \cdot x_1^{p_1,1} \ldots {x_{k-1}}^{p_{k-1,1}} + \ldots +  a_t \cdot x_1^{p_1,t} \ldots {x_{k-1}}^{p_{k-1,t}}
\end{equation}
for some $t$, where $a_i$ are unknown constant coefficients and $p$ are exponents taken from the skeleton.

This is a linear system for $a_i$ and hence knowing the values of $f(x_1, \ldots, x_{k-1}, c_i, \ldots)$ for $t$ different sets of $\{x_1, \ldots, x_{k-1}\}$ (sampling points or probes) we can solve the system. This however has a complexity $O(t^3)$, thus the Zippel algorithm for polynomials suggests a specific set of sampling points, i.e.\
${y_1, ... y_{k-1}}, {y_1^2, ... y_{k-1}^2}, \ldots, {y_1^t, ... y_{k-1}^t}$. In this case the system turns into a Vandermonde system which can be solved with complexity $O(t^2)$.
The Zippel method consists of solving this system leading to the knowledge of $f(x_1, \ldots, x_{k-1}, c_i, \ldots)$ for different $i$, followed by univariate Newton reconstruction in $x_k$ to obtain $f(x_1, \ldots, x_{k-1}, x_k, \ldots)$.

For the tasks described one needs to reconstruct rational functions of multiple variables, but it is possible to adapt the Zippel approach, for example, as the balanced Zippel method described in \cite{Smirnov:2024onl}.

What is also important for the proper reconstruction order is to perform it in modular arithmetic over large prime numbers. This prevents coefficient growth and decreases the number of needed sampling points, and the final reconstruction to rational numbers is performed as a final step. Thus to proceed with the algorithms one needs an efficient library working with modular polynomials of multiple variables. There is not a large number of such libraries, \texttt{FLINT} \cite{Hart2010, FLINT} being one of the most efficient. 

\section{The proposed GPU Implementation of the Zippel Method}
\label{sec:impl}

Most nodes of modern supercomputers and clusters are now equipped with dedicated GPUs that enable parallel execution of mathematical operations on large datasets in a multithreaded mode. When implemented correctly, they can significantly reduce the computation time during the program execution.

Modern GPUs are capable of handling thousands of threads simultaneously, offer excellent horizontal scalability, and their architecture is optimized for the simultaneous execution of uniform operations — a critical feature for the scientific computing tasks performed by the \texttt{FIRE} program. 

For the case of multiple variables the Zippel reconstruction step sometimes becomes a bottleneck for the whole reduction process, hence implementing it on a GPU seemed an important task.

\subsection{Modular integer operations on GPUs}
    
There are not many ready-to-use solutions for modular arithmetic on GPU. Only after implementing our solution we found a \texttt{CUMODP} (CUDA Modular Polynomial Library) library.
A further investigation showed that this library uses 32-bit modular integers so is not directly ready for our case, which requires 64-bit operations. 

Thus during the development of the GPU migration solution we tested several approaches:
\begin{itemize}
    \item standard modular arithmetic operators provided by the \texttt{nvcc} compiler using the \texttt{uint128\_t} data type;
    \item an algorithm leveraging properties of modular arithmetic with the \texttt{uint64\_t} data type with special GPU intrinsics;
    \item a ported version from \texttt{CUMODP} to the 64-bit case;
    \item and GPU-ported basic functions from the \texttt{FLINT} library (which is used in the original \texttt{FIRE} application).
\end{itemize}

Let us describe those variants in details and compare their efficiency on a multiplication modular some large prime $n$ fitting into \texttt{uint64\_t}.

Variant 1 is the most straightforwad using the \texttt{uint128\_t} type, here we completely rely on the compiler:

    \begin{lstlisting}[language=C++]
    // return (a * b) mod n
    __inline__ __device__
     unsigned long long mul_mod(
     unsigned long long a, 
     unsigned long long b, 
     unsigned long long n)
    {
        return (static_cast<__uint128_t>(a) * b ) % n;
    }    
    \end{lstlisting}

A better approach (number 2) is to avoid the \texttt{uint128\_t} type stating in \texttt{uint64\_t} and using special CUDA intrinsic, \texttt{\_\_umul64hi(a, b)} which takes two 64-bit numbers and as a result returns the high part of its multiplication. This approach lets one avoid using the \texttt{uint128\_t} type. The code can be implemented the following way:

     \begin{lstlisting}[language=C++]
    __inline__ __device__
    unsigned long long mul_mod( 
        unsigned long long a, 
        unsigned long long b, 
        unsigned long long n)
    {
        // Calculate the most significant 64 bits 
        // of the product of the two 64 unsigned bit integers.
        unsigned long long hi = __umul64hi(a, b); 
        unsigned long long lo = a * b;
        hi %= n;
        lo %= n;
        // pow = 2^64 mod n
        unsigned long long pow = (1ULL << 63) % n;
        pow = ( pow << 1 ) % n;
        hi = ( hi * pow ) % n;

        return (hi + lo) % n;
    }
     \end{lstlisting}

Variant 3 is based on \texttt{CUMODP} and uses a fast modular multiplication algorithm based on floating-point arithmetic. It has several significant limitations, the main one being its dependence on floating-point precision -- the mantissa must be large enough to hold the full range of values without rounding errors. In this example, it has been extended to support larger numbers by using quadruple-precision floating-point numbers. The code looks the following way:

\begin{lstlisting}[language=C++]
__inline__ __device__
int64_t mul_mod1(int64_t a, int64_t b, int64_t n) {
    long double ninv = 1.0L / (long double) n;
    int64_t q = (int64_t)((((long double)a) * ((long double)b)) * ninv);
    int64_t res = a * b -  q * n;
    res += (res >> 63) & n;
    res -= n;
    res += (res >> 63) & n;
    return res;
}
\end{lstlisting}

The \texttt{FLINT} library is open-source, written in C, and optimized for 32/64-bit arithmetic, which allows its code to be ported to \texttt{CUDA} with minimal modifications. However the implementation is large enough due to many functions that are required. The top-layer function for multiplication looks as follows:
     \begin{lstlisting}[language=C++]
     __device__  
     mp_limb_t nmod_mul_d(mp_limb_t a, mp_limb_t b, nmod_t mod)
     { 
        b <<= mod.norm;
        mp_limb_t res;
        do {
            mp_limb_t q0xx, q1xx, rxx, p_hixx, p_loxx;
            mp_limb_t nxx, ninvxx;
            unsigned int normxx;
            ninvxx = (mod).ninv;
            normxx = (mod).norm;
            nxx = (mod).n << normxx;
            umul_ppmm_d(p_hixx, p_loxx, (a), (b));
            umul_ppmm_d(q1xx, q0xx, ninvxx, p_hixx);
            add_ssaaaa_d(q1xx, q0xx, q1xx, q0xx, p_hixx, p_loxx);
            rxx = (p_loxx - (q1xx + 1) * nxx);
            if (rxx > q0xx) rxx += nxx;
            rxx = (rxx < nxx ? rxx : rxx - nxx) >> normxx; 
            (res) = rxx;
        } while (0);
        return res;
     }
     \end{lstlisting}

We ran a benchmark for all variants and got the following results. The multiplication was performed modulo two numbers located in global memory. The tests were performed on Radeon RX 560X (163 GFlops). The test involved 1 core and 1 thread.

\begin{table}[h!]
\centering
\caption{Results of additional fuzzing tests for modular multiplication (10,000,000 iterations)}
\begin{tabular}{|l|c|}
\hline
\textbf{Algorithm} & \textbf{Time (sec)} \\
\hline
\texttt{uint128\_t} & 140.13 \\
\texttt{uint64\_t}  & 24.67 \\
\texttt{CUDMODP*}   & 9.96 \\
\texttt{FLINT}      & 1.48 \\
\hline
\end{tabular}
\label{tab:1}
\end{table}

As we can see from the table~\ref{tab:1}, both native implementations are quite inefficient. The reason is that the GPUs themselves are 32-bit machines, so even the 64-bit operations are supported but are not natural for the cores implemented as consequent 32-bit operations. The \texttt{CUMODP} implementation also has problems due to a division in the code. The \texttt{FLINT} version is an obvious winner.

Thus we decided to port the \texttt{FLINT} multiplication to GPUs using all its powerful features including the precalculated inversion letting one to avoid the division operations. The following functions were ported for modular addition, multiplication, exponentiation, inversion, and remainder operations for 128- and 192-bit composite integers:
\texttt{nmod\_mul}, \texttt{add\_ssaaaa}, \texttt{umul\_ppmm}, \texttt{NMOD\_RED2}, \texttt{NMOD\_RED3}, \texttt{n\_gcdinv}, \texttt{nmod\_pow}, \texttt{n\_invmod}, \texttt{nmod\_inv}. We are considering to later provide the ported code as a standalone library.

\subsection{Zippel algorithm implementation details}

The next step was to benchmark the CPU implementation of the Zippel algorithm in order to determine the algorithm parts requiring optimization. This was performed both with profilers and by manual code analysis searching for parts with quadratic complexity by the number of terms $t$ in the skeleton monomial. 

The most resource-intensive parts of the algorithm are two functions: \texttt{ZippelMultiplePrime()}, which performs a Zippel reconstruction of several expressions of various degrees in the simple case, and \texttt{BalancedZippel()}, which prepares a balanced numerator and denominator for a subsequent call to \texttt{ZippelMultiplePrime()}.

The \texttt{ZippelMultiplePrime()} function computes values of expressions at specified points using Lagrange interpolation. The estimated theoretical complexity for a single thread is high; however, with efficient parallelization, the complexity can be reduced to $O((n^2 + n m)/k)$, where $k$ is the number of threads. The main steps of the Zippel algorithm can be illustrated as follows:

\begin{itemize}
    \item Polynomial values at specified points are computed using Horner's scheme.
    \item During reconstruction, a vector of powers is generated in parallel.
    \item This vector of powers is multiplied by the vector of polynomial values.
    \item The resulting vector is scalar-multiplied by the modular inverse of the polynomial value computed via Horner's method.
    \item The final result vector is written to memory.
\end{itemize}

Several optimizations were introduced during GPU porting.

First, we used aligned memory, which plays a crucial role in working with two-dimensional data arrays on GPUs. Data is arranged in memory such that the addresses are powers-of-two aligned, with the alignment degree depending on the GPU architecture. This approach minimizes memory access latency and increases bandwidth during parallel access to array elements.

Second, the power vector in this algorithm is temporary and can be replaced with an accumulated sum whose length does not exceed~$t$. Since access to this accumulator is frequent, placing it in shared memory provides the best data access performance.

Third, the modular addition operation which normally requires a modulo reduction at each step can be replaced by a standard summation with overflow control (emulating a 192-bit integer). This allows deferring a costly modulo operation until the final data processing stage.

\[
(...(((a_1 \cdot b_1) \bmod N + (a_2 \cdot b_2) \bmod N) \bmod N + \dots) \bmod N) \equiv 
((a_1 \cdot b_1) + (a_2 \cdot b_2) + \dots) \bmod N
\]

The schematic representation of summing a large number of 64-bit values followed by modulo reduction is as follows:

\begin{lstlisting}[language=C++]
...
ulong tempAccumLow = 0, TempAccumMid = 0, TempAccumHi = 0;
... 
for (...) {
    ...
    // Multiply two 64-bit integers, result being split
    // into high (p1) and low (p0) parts
    umul_ppmm(p1, p0, term, value);

    // Add p1 and p0 to the accumulator
    add_ssaaaa(tempAccumMid, tempAccumLow, tempAccumMid, tempAccumLow, 
        p1, p0);

    // Overflow control
    if (tempAccumMid < p1) tempAccumHi++;
    ...
}
...
// Equivalent to: ((a << 128) | (b << 64) | c) % N
NMOD_RED3(res, TempAccumHi, tempAccumMid, tempAccumLow, N);
...
\end{lstlisting}

Replacing the temporary power vector in the \texttt{ZippelMultiplePrime()} function also significantly reduces the amount of memory consumed on the GPU, which is critical when processing data in a large number of parallel threads.
For example, processing a polynomial of approximately 5 million terms would require an additional 0.3 GB of memory per thread to store intermediate results.
   
Another block ported to the GPU is the preparation of balanced coefficients.
The balancing approach involves constructing expressions by simultaneously multiplying both the numerator and denominator by an additional numerical factor. This allows the numerator and denominator to be independently reconstructed using the Zippel algorithm (for details see \cite{Smirnov:2024onl}).

\begin{eqnarray}
\frac{n_i(y_1^i, \ldots, y_{k-1}^i, c_0^j, \ldots) \cdot n'(y_1^i, \ldots, y_{k-1}^i, c_0, \ldots)}
{n_i(y_1^i, \ldots, y_{k-1}^i, c_0, \ldots)}  &=& \nonumber \\ 
\frac{n(y_1^i, \ldots, y_{k-1}^i, c_0^j, \ldots) \cdot h_i \cdot n(y_1^i, \ldots, y_{k-1}^i, c_0, \ldots) \cdot C}
{n(y_1^i, \ldots, y_{k-1}^i, c_0, \ldots) \cdot h_i} &=&
n(y_1^i, \ldots, y_{k-1}^i, c_0^j, \ldots) \cdot C.
\end{eqnarray}

To construct such expressions, it is necessary to raise base variable values to various powers and substitute them into the polynomial on the GPU.

For this purpose, a fast exponentiation algorithm is used. A two-dimensional array of coefficients is generated by multiplying a base value by a precomputed $k$-th power of the same value, where $k$ is the width of the array. This approach allows for quickly obtaining a coefficient array with a step of $k$:

\begin{eqnarray}
{ x_1^n, x_1^{n-1}, \ldots, x_1^{n-k} } &=& x_1^k \cdot { x_1^{n-k}, x_1^{n-k-1}, \ldots, x_1^{n-2k} } \nonumber \\
{ x_2^n, x_2^{n-1}, \ldots, x_2^{n-k} } &=& x_2^k \cdot { x_2^{n-k}, x_2^{n-k-1}, \ldots, x_2^{n-2k} } \nonumber \\
\vdots \nonumber \\
{ x_m^n, x_m^{n-1}, \ldots, x_m^{n-k} } &=& x_m^k \cdot { x_m^{n-k}, x_m^{n-k-1}, \ldots, x_m^{n-2k} } \nonumber
\end{eqnarray}

Each GPU thread (kernel) processes its own value independently of others. Second-level loops are parallelized inside the kernel using individual threads. Each thread computes a portion of the polynomial.
The final polynomial values are then summed together.

GPUs offer flexible mechanisms for inter-thread communication within the same warp, which provides significantly faster data exchange compared to global or shared memory access.
In this case, inter-thread communication is implemented using the built-in function \texttt{\_\_shfl\_down\_sync()}, which allows threads within a warp to exchange data in a downward direction.

Below is an example of a \texttt{warpReduceSum()} function, which sums all values across threads within a warp without relying on global or shared memory:

\begin{lstlisting}[language=C++]
__inline__ __device__ 
mp_limb_t warpReduceSum(ulong val, nmod_t flint_mod) 
{
    for (int shift = warpSize/2; shift > 0; shift /= 2)
    {
        val = nmod_add_d(val, 
                        __shfl_down_sync(warpSize - 1, val, shift), 
                        flint_mod);
    }
    return val;
}
\end{lstlisting}

To conclude, the implemented algorithm enables offloading the most resource-intensive part — coefficient reconstruction — to the GPU, thereby providing a flexible horizontal scalability. It also means that this implementation can be easily adapted for running on multiple devices or cluster nodes since in real physical examples there are multiple coefficients which have to be reconstructed, so that they can be discributed among different GPUs and even supercomputer nodes. Additionally, this approach frees up CPU resources, which can then be used for other tasks. The resulting performance will be discussed in the next section.

\section{Performance Evaluation and Analysis}
\label{sec:eval}

After developing the proposed GPU solution, it was necessary to evaluate its performance and speedup in comparison with the original CPU version (which is available as a private version of \texttt{FIRE} but with possible pre-publication access by request).

\subsection{Experimental Conditions}

Experiments were carried out on three GPUs --- NVIDIA V100, P100 and A100, and Intel Xeon Gold 6126 2.6 GHz (12 cores, 178 Gop/s peak for INT64 operations) was used for comparing with the original CPU version. The main characteristics of used GPUs are shown in Tab.~\ref{tab:gpus}. The main type of data used is int64, so the theoretical peak performance was calculated for this operation type.
Experiments were conducted on the equipment of the Supercomputer Center of Lomonosov Moscow State University: ``pascal'' partition of the Lomonosov-2 supercomputer~\cite{voevodin2019lomonosov2} was used for tests on P100; ``volta1'' Lomonosov-2 partition --- for tests on V100 and Intel Xeon 6126; calculations on A100 were performed on a standalone server.

\tab{tab:gpus}{Characteristics of GPUs used for performance evaluation}{
	\begin{tabular}{ | c | c | c | c | }
		\hline
        \textbf{GPU name} & P100 & V100 & A100 \\
        \hline 
    	\textbf{INT64 peak performance, Top/s} & 2.38 & 3.53 & 4.87 \\
        \hline 
        \textbf{Number of CUDA cores} & 3584 & 5120 & 6912 \\
        \hline 
        \textbf{Theoretical peak memory bandwidth, GB/s} & 720 & 900 & 1555 \\
        \hline 
        \textbf{Memory size, GB} & 16 & 32 & 40 \\
        \hline 
	\end{tabular}
}

Four different input data sizes were considered, all related to the final reconstruction over the sixth variable ($d$, the space-time dimension) using the Zippel algorithm. The datasets correspond to the same physical problem and represent reductions to master integrals of varying level. As a result, they differ in the number of monomials in the skeleton polynomial. The selected examples include the following numbers of monomials: 125 thousand (``125k''), 300 thousand (``300k''), 750 thousand (``750k''), and 4.8 million (``4.8m''). The computational complexity (and therefore the calculation time) does not depend much on the structure of the datasets, only on their size, so there was no need to consider different dataset variants.

As mentioned earlier, the complexity of the Zippel algorithm grows quadratically with the number of monomials, which makes this dataset sufficiently representative. In particular, the last example was barely within the time constraints when computed on a CPU on the Lomonosov-2 supercomputer.

Each test was repeated 10 times in order to collect enough statistics on statistical significance between execution time of different experiments (the only exception is that for ``4.8m'' size the number of launches on CPU was reduced to 5, as they require a lot of time to run). The differences in execution times between identical runs were minimal and there were no outliers, so only average values are reported below. 

\subsection{Evaluations Results}

Tab.~\ref{tab:comparison1} shows the results of execution time comparison on four platforms and for four input data sizes as described above. The top row shows the execution time in seconds for CPU version on Intel Xeon, while the rows below show the speedup on GPUs compared to Intel Xeon result.

\tab{tab:comparison1}{Comparison of execution time for different platforms}{
	\begin{tabular}{ | c | c | c | c | c | }
        \hline
		   & \multicolumn{4}{c|}{\textbf{Input data size}} \\
		\hline
		\textbf{Platform} & \textbf{125k} & \textbf{300k} & \textbf{750k} & \textbf{4.8m} \\
		\hline 
		Xeon 6126, time & 44 s & 255 s & 1550 s & 74431 s \\
		\hline 
        \hline 
        P100, speedup & 2.69 & 3.23 & 3.55 & 3.84 \\
        \hline 
        V100, speedup & 3.87 & 5.36 & 7.69 & 8.85 \\
        \hline 
	   	A100, speedup & 3.81 & 7.46 & 10.76 & 14.57 \\
        \hline 
	\end{tabular}
}

We can see that execution time on Intel CPU varies from less than 1 minute up to 20+ hours, showing a wide duration spectrum. On ``125k'' input size, GPU speedup is up to 4, and it starts to increase as the input size gets bigger, leading up to 14.5x speedup on A100 on ``4.8m'' input size. It is interesting to note that on ``125k'' input size the speedup for V100 is shown to be bigger than the speedup for A100, although the latter GPU is more modern and powerful, but this difference is actually statistically insignificant (confidence intervals are overlapping). The difference between all other results is statistically significant.

In Tab.~\ref{tab:comparison2}, the same results for GPUs are presented, but the speedup on the V100 and A100 platforms relative to the P100 is shown, which allows for an easier comparison of execution time difference between three GPUs. We can see that on ``125k'' input size, the speedup is not so big --- less than 1.5 times both for V100 and A100, but on ``4.8m'' input size it rises up to 2.3 for V100 and 3.8 for A100 GPU. Thus, for small task sizes, the use of more powerful graphics accelerators does not provide a significant benefit, but as the size increases, the advantage of more powerful GPUs becomes more and more noticeable.

\tab{tab:comparison2}{Comparison of execution time between different GPUs}{
	\begin{tabular}{ | c | c | c | c | c | }
        \hline
		   & \multicolumn{4}{c|}{\textbf{Input data size}} \\
		\hline
		\textbf{Platform} & \textbf{125k} & \textbf{300k} & \textbf{750k} & \textbf{4.8m} \\
		\hline 
        Speedup of V100 compared to P100 & 1.44 & 1.66 & 2.17 & 2.31 \\
        \hline 
        Speedup of A100 compared to P100 & 1.42 & 2.31 & 3.03 & 3.80 \\
        \hline 
	\end{tabular}
}

A few words about the efficiency of the proposed implementation. For its rough estimation, the GPU load metric was used, which shows the percent of time during which one or more kernels was active (executing on the GPU). According to preliminary information (obtained not for all types of experiments), the efficiency grows with the increase of the task size, and for the case of ``4.8m'' it reaches values above 90\% on all three types of GPU. For the size ``700k'' this value is on average not lower than 80\%. A more detailed analysis is planned in the future, but it can already be concluded that the efficiency of the obtained implementation is high. It is also worth noting that the utilization within the CPU version is also high --- a commonly used CPU load metric (percentage of time spent in user program) shows on average a value above 95\%.

\section*{Conclusion}
\label{sec:conclusion}
In this paper, a GPU implementation of the Zippel method based on porting and adapting \texttt{FLINT} functions is proposed. According to our information, this is the first implementation of this algorithm on GPUs. It provides a significant speedup when compared to a CPU variant. We intend to make the code public this year as a part of the new \texttt{FIRE} version and expect this implementation to be used for different reduction tasks in elementary particle physics, both as part of the \texttt{FIRE} package and without it, since the reconstruction utilities can be used stand-alone. It is in our plans to speed up the reconstruction even more with the use of texture memory.

\section*{Acknowledgements}
The work was supported by the Ministry of Education and Science of the Russian Federation as part of the program of the Moscow Center for Fundamental and Applied Mathematics under Agreement No. 075-15-2025-345. The research was carried out using the equipment of the shared research facilities of HPC computing resources at Lomonosov Moscow State University \cite{voevodin2019lomonosov2}.

\openaccess

\bibliography{example}
\end{document}